\documentclass[english,aps,prb,amsmath,amssymb,floatfix,superscriptaddress,eqsecnum,longbibliography,twocolumn]{revtex4-2}
\usepackage[T1]{fontenc}
\usepackage{tabularx}
\usepackage{booktabs}
\usepackage{siunitx}
\usepackage{amsmath}
\usepackage{graphicx}
\usepackage{txfonts}
\usepackage{Jonasmacros}
\usepackage{xcolor}
\usepackage{ulem}
\usepackage[unicode=true,pdfusetitle,
 bookmarks=true,bookmarksnumbered=false,bookmarksopen=false,
 breaklinks=false,pdfborder={0 0 1},backref=false,colorlinks=false]
 {hyperref}
\begin{document}

\title{ 
%nonrelativistic Dzyaloshinskii-Moriya interactions in the coplanar antiferromagnet Mn$_3$Sn \\ or \\
Unraveling the connection between high-order magnetic interactions and local-to-global spin Hamiltonian in noncollinear magnetic dimers
 %\\ or \\
% The Dzyaloshinskii-Moriya interaction in Mn$_3$Sn is of non relativistic origin
}

% possible alternatives:

%Large Dzyaloshinskii-Moriya interaction in absence relativistic effects

%The Dzyaloshinskii-Moriya interaction in Mn$_3$Sn is of non relativistic origin

\author{Ramon Cardias}
\affiliation{Department of Applied Physics, School of Engineering Sciences, KTH Royal Institute of Technology, AlbaNova University Center, SE-10691 Stockholm, Sweden}
\affiliation{Instituto de Física, Universidade Federal Fluminense, 24210-346, Niterói RJ, Brazil}

\author{Jhonatan dos Santos Silva}
\affiliation{Faculdade de F\'\i sica, Universidade Federal do Par\'a, Bel\'em, PA, Brazil}
\affiliation{Instituto Federal de Educa\c{c}\~ao, Ci\^encia e Tecnologia do Par\'a, \'Obidos, PA, Brazil}

\author{Anders Bergman}
\author{Attila Szilva}
\author{Yaroslav O. Kvashnin}
\author{Jonas Fransson}
\affiliation{Department of Physics and Astronomy, Uppsala University, 75120 Box 516 Sweden}

\author{Angela B. Klautau}
\affiliation{Faculdade de F\'\i sica, Universidade Federal do Par\'a, Bel\'em, PA, Brazil}
\affiliation{Departamento de F\'\i sica da Universidade de Aveiro, 3810-183 Aveiro, Portugal}

\author{Olle Eriksson}
\affiliation{Department of Physics and Astronomy, Uppsala University, 75120 Box 516 Sweden}
\affiliation{Wallenberg Initiative Materials Science for Sustainability (WISE), Uppsala University, 75120 Box 516, Sweden}

\author{Anna Delin}
\affiliation{Department of Applied Physics, School of Engineering Sciences, KTH Royal Institute of Technology, AlbaNova University Center, SE-10691 Stockholm, Sweden}
\affiliation{Wallenberg Initiative Materials Science for Sustainability (WISE), KTH Royal Institute of Technology, SE-10044 Stockholm, Sweden}
\affiliation{SeRC (Swedish e-Science Research Center), KTH Royal Institute of Technology, SE-10044 Stockholm, Sweden}

\author{Lars Nordstr\"om}
\affiliation{Department of Physics and Astronomy, Uppsala University, 75120 Box 516 Sweden}

\date{\today}

\begin{abstract}

A spin Hamiltonian that characterizes interatomic interactions between spin moments is highly valuable in predicting and comprehending the magnetic properties of materials. 
Here, we explore a method for explicitly calculating interatomic exchange interactions in noncollinear configurations of magnetic materials considering only a bilinear spin Hamiltonian in a local scenario. 
Based on density-functional theory (DFT) calculations of dimers adsorbed on metallic surfaces, and with a focus on the Dzyaloshinskii-Moriya interaction (DMI) which is essential for stabilizing chiral noncollinear magnetic states, we discuss the interpretation of the DMI when decomposed into microscopic electron and spin densities and currents.
We clarify the distinct origins of spin currents induced in the system and their connection to the DMI. 
%%%%%%%%
%%%%%%%
In addition, we reveal how noncollinearity affects the usual DMI, which is solely induced by spin-orbit coupling, and DMI-like interactions brought about by noncollinearity. 
We explain how the dependence of the DMI on the magnetic configuration establishes a connection between high-order magnetic interactions, enabling the transition from a local to a global spin Hamiltonian. 
%PACS numbers: 75.75.+a, 73.22.-f, 75.10.-b
\end{abstract}

\maketitle

\section{Introduction}

In order to calculate the magnetic properties of a given system or material, one of the most used approaches is to start from a spin Hamiltonian that describes the interatomic interactions between the atomic spins. This approach -- the atomistic approach -- assumes that it is possible to identify well-defined regions in the material where the magnetisation density is more or less unidirectional and sizeable only close to an atomic nucleus. One of the best examples of a spin Hamiltonian is the generalized bilinear classical
spin model, which is written as
\begin{align}
	\mathcal{H}&=-\sum_{\left<ij\right>}\left(J_{ij}\vec{e}_i{\cdot} \vec{e}_j+\vec{D}_{ij}{\cdot}\left(\vec{e}_{i}{\times}\vec{e}_{j}\right)+\vec{e}_i{\cdot}\mathcal{A}_{ij}{\cdot} \vec{e}_j\right), 
	\label{ham}
\end{align}
%\AnnaD{Are you sure you do not prefer minus signs in the equation above?}
where $J_{ij}$ is the scalar Heisenberg exchange coupling parameter between the spins at atoms $i$ and $j$, $\vec{D}_{ij}$ is the Dzyaloshinskii-Moriya  interaction (DMI) vector, $\mathcal{A}_{ij}$ is the symmetric anisotropic interaction term, $\vec{e}_i$ and $\vec{e}_j$ are unit vectors describing the directions of the atomistic spin moments at site $i$ and $j$, respectively. The summation is made over pairs of atoms $\left<ij\right>$. Many methods have been proposed in the recent literature regarding how to calculate these parameters \cite{yang2023first,szilva_quantitative_2022,Jacobsson1715603,PhysRevB.101.174401,PhysRevB.104.054418}. A widely used such method is the one proposed in Refs.~\cite{liechtenstein_exchange_1984,liechtenstein_local_1987} and recently reviewed in Ref.~\cite{szilva_quantitative_2022}. We will refer to this approach as the Liechtenstein-Katsnelson-Antropov-Gubanov (LKAG) method. It is based on the magnetic force theorem~\cite{liechtenstein_exchange_1984,liechtenstein_local_1985}, which assumes that an effective spin Hamiltonian accurately describes the energy landscape of the atomic spin configurations sufficiently close to the magnetic ground state. In other words, the electronic Kohn-Sham Hamiltonian can be mapped onto a (classical) spin Hamiltonian since the variation of the total energy of the electronic subsystem can be expressed in terms of variations only of occupied single particle states.

It is known that in all LKAG-like approaches  
%, especially when the magnetic state is allowed to have any noncollinear order~\cite{szilva_interatomic_2013},
the interatomic exchange coupling parameters depend on the underlying magnetic state from which they are calculated. This becomes especially clear when the magnetic state is allowed to have noncollinear order, %~\cite{szilva_interatomic_2013}
since then this magnetic state can be varied continuously and the corresponding variation of the calculated bilinear interatomic exchange coupling parameters is obvious~\cite{cardias_dzyaloshinskii-moriya_2020}. Another way to view this is to say that the mapped spin model is local, which means that it is only valid 
%in a local region of the energy vs. configuration curve, i.e., it is only relevant 
for small magnetic variations around the reference magnetic state. 

This is in contrast to the concept of a global spin model, which is valid for all possible spin configurations.
It has been argued that such a global model is possible if it also incorporates higher order spin interactions beyond bilinear couplings, i.e., multiple interactions.  It has been shown that such terms appear naturally in a perturbative approach where the reference state is nonmagnetic, in contrast to the magnetic reference state in LKAG \cite{Brinker_2019}. However, a disadvantage with such an approach is that the perturbation from the reference state can be considerable. In contrast, the LKAG approach leads to a slow convergence in terms of multi-spin interactions.
%In practice, this means that one needs to consider electron hopping, not only between the two crystal atomic sites considered, but also between other sites and one can think of
%these high-order terms as multi-spin interactions
~\cite{streib_adiabatic_2022,streib_exchange_2021,cardias_comment_2022,brinker_prospecting_2020}. 
The convergence of a multispin model eventually becomes a combinatorial problem, and it becomes very difficult to reach completeness, i.e.~to determine how many multi-spin interactions should be considered. With increasing complexity of the multi-spin interactions, it becomes hard to get a clear grasp of their origin and physical meaning.

In this study we have chosen to simplify the magnetic structure as far as possible.
Even in such a simple system as a magnetic dimer, with a varying noncollinear magnetic configuration, the bilinear interactions are heavily reference-state dependent and the corresponding spin model is hence nontrivial ~\cite{cardias_spin_2021}. In the mapping to a spin model for magnetic dimers,  two complementary approaches will be compared. Firstly, the reference state is taken explicitly into account in an LKAG-like approach and the magnetic interactions are calculated as a function of the magnetic configuration. Secondly, we will recast the bilinear interactions as a reference-state independent spin model with higher order interactions, by means of sum rules of the Green function. In this way, we can illustrate the connection between the magnetic configuration-dependent and multi-spin representations in the case of dimers, by explicitly showing how the DMI interaction behaves in the two pictures, and simultaneously provide insight into its microscopic origins.

More specifically, we have developed a technique to calculate the interactions described in Eq.~\ref{ham} for any magnetic configuration considering only a bilinear spin Hamiltonian~\cite{cardias_dzyaloshinskii-moriya_2020,cardias_first-principles_2020}. 
%%%%%%%%%%%%%%%
%It has been revealed the magnetic moment dependence of these parameters where it can be interpreted as the emergence of high-order terms folded in our bilinear Hamiltonian terms as the noncollinear magnetic texture arises. 
%%%%%%%%%%
The dependence of these parameters on the magnetic configuration has been revealed, which can be interpreted as the emergence of high-order terms folded onto a bilinear Hamiltonian expression as the noncollinear magnetic texture arises.
%%%%%%%%%
In this way, one does not have access to the multi-spin perspective of the problem, but on the other hand, one can consider a simple solution to study the magnetic properties locally in configuration space. This approach (technique), combined with atomistic spin dynamics, has been shown to significantly refine the comparison between theory and experiment when considering properties of excited states, such as the magnon softening induced by temperature effects, as demonstrated in Refs.~\onlinecite{szilva_interatomic_2013} and ~\onlinecite{rodrigues_finite-temperature_2016}. 
Moreover,
in Refs.~\onlinecite{cardias_first-principles_2020} and \onlinecite{cardias_dzyaloshinskii-moriya_2020} we have shown that these interactions can be seen in terms of spin/charge density and spin/charge currents. 
%{\color{red}
It was recently demonstrated that the DMI calculated from a noncollinear magnetic configuration can have a large magnitude\cite{cardias_first-principles_2020,cardias_dzyaloshinskii-moriya_2020} and, most importantly, have a nonchiral behaviour, which appears to contradict the original works of Moriya and Dzyaloshinskii~\cite{moriya_anisotropic_1960,dzyaloshinsky_thermodynamic_1958}. This observation has contributed to a vivid discussion~\cite{dos_santos_dias_proper_2021,cardias_comment_2022}, about the mechanisms behind these interactions and their possible equivalence to high-order terms and/or multi-spin interactions in a global model. 
A more extensive discussion on spin Hamiltonians and how to correctly map them from an electronic Hamiltonian can be found in Ref.~\cite{szilva_quantitative_2022} and references therein.
%{\bf Ramon, I have a feeling that this claim goes back long before the works in Refs. 15,16 and 20. Can we add some older reference here instead ? Maybe the first LKAG paper or even older stuff, or if we are Lazy we can just add a referemce to Attilas review. At least it is a larger discussion on spin-Hamiltonians.

%Ramon: I opted for the lazy option... but I also think it is efficient in that case.}
%Nevertheless, they were unable to explain the strong values found by us. 

This paper is organized as follows: In Sec. \ref{sec2}, we discuss the DMI structure -- determined through first-principles calculations -- and its relation to spin and charge currents. In Sec.\ref{sec3}.A, we use the RS-LMTO-ASA method~\cite{klautau_magnetic_2004,klautau_orbital_2005,bergman_magnetic_2006,bergman_magnetic_2007,bergman_non-collinear_2006} to ascertain the magnetic states and interactions for each system addressed. Sec.\ref{sec3}.B explores the impact of structural relaxations on the current study's conclusions. Sec.\ref{sec3}.C examines the origins of spin currents and their association with the DMI. Sec.\ref{sec3}.D presents the DMI between two spin moments across various magnetic configurations, with a focus on the effect of noncollinearity on the DMI. Finally, in Sec.\ref{sec3}.E, we delve into the magnetic dependence of the DMI, highlighting its connection with high-order magnetic interactions and the transition from local to global spin Hamiltonians.

\section{Method}
\label{sec2}

The Kohn-Sham equation, which has the form of a single particle Schr\"{o}dinger equation, can be written as
\begin{eqnarray}
\label{schroed}
i\frac{\partial \psi }{\partial t} &=&H\psi  \nonumber \\
H &=&-\nabla ^{2}+V({\bf r})- \left(\vec{B}_{xc}({\bf r})+{\vec{B}}_{ext}( {\bf r}) \right) \cdot \vec{\sigma} \,
\label{firstKS}
\end{eqnarray}
where $V({\bf r})$ is the effective
potential, ${\vec{B}}_{ext}({\bf r})$ and ${\vec{B}}_{xc}({\bf r})$
are the external magnetic field and the
exchange-correlation field, respectively, which couple to the electrons spin, and $\vec{\sigma}$ stands for the Pauli spin matrices $\{\sigma_x,\sigma_y,\sigma_z\}$. Note that Rydberg units are used here: $\hbar=2m=e^2/2=1$ (it is also noteworthy that in Appendix~\ref{app:A}, we use a different definition of the wave function, using capital $\Psi$).  From the Kohn-Sham Hamiltonian, one can calculate the Green function as 
\begin{equation}
G(z)=\left(z-H\right)^{-1}\;, 
\label{five}
\end{equation}
where $z \in \mathbb{C}$. Note that $G(z)$ can be decomposed into intersite terms, $G_{ij}$, since $G(z)=\sum_{ij} |\phi_{i} \rangle G_{ij} \langle \phi_{j}|$ with local functions $|\phi_i\rangle$ at site $i$, which can be further decomposed into spin-components as
\begin{align}
	G_{ij}=G^0_{ij}+\vec{G}_{ij}\cdot\vec{\sigma}\,,\label{GF-s}
\end{align}
where $\vec{G}_{ij}$ is a vector with the components of $G^{\eta}_{ij}$ where the index $\eta$ enumerates both the scalar spin-independent Green function as well as the components of the spin-dependent vector Green function of Eq.~\eqref{GF-s}, i.e.~$\eta$ can be either $0$, $x$, $y$ or $z$. We will refer here to  $G^0_{ij}$ as the charge part and to $\vec{G}_{ij}$ as the spin part of the Green function. Note that one can further decompose the components of the Green function into terms that are either even or odd under time-reversal symmetry \cite{fransson_microscopic_2017}. This can be done by introducing $G_{ij}^{\eta \kappa}$ where the second index $\kappa$ can be viewed as an indicator whether the terms that are time reversal invariant and those are not, i.e, $\kappa$ can be $0$ or $1$ (the exact relation is explained below). This decomposition of the Green function can be summarised as,
\begin{align}
G^\eta_{ij}&=G^{\eta 0}_{ij}+G^{\eta 1}_{ij}, \nonumber\\
\label{GF-st}
\end{align}
where $G^{00}_{ij}$ and $\vec{G}^1_{ij}$  are time reversal invariant while $G^{01}_{ij}$ and $\vec{G}^0_{ij}$ are not. Sometimes it is convenient to write the $x$, $y$, or $z$ components of the Green function as vectors, i.e.~$\vec{G}^\kappa$. This decomposition also plays a useful role in how the Green function behaves under site exchange, since in a real local basis \cite{fransson_microscopic_2017} we have that
\begin{align}
    G^{\eta \kappa}_{ij}=(-1)^{\kappa}{G^{\eta\kappa}_{ji}}^T \,.
    \label{etakappatrafo}
\end{align}
In fact, it has been shown that these two index Green functions are decomposed in terms that produce local charge-, $G^{00}$, or spin-densities, $\vec{G}^{0}$, and charge-, $G^{01}$, and spin-currents $\vec{G}^{1}$, respectively \cite{szilva_quantitative_2022}. Once the Green function is given, the (integrated) density of states and the grand canonical potential ($\Omega$) of the electronic sub-system can also be determined.

As a next step, and as reviewed in Ref.\cite{szilva_quantitative_2022}, one can introduce a small variation of the atomic spin at site $i$ and derive the variation of the grand potential compared to a reference state (the ground state). The same procedure can be done at the level of a spin Hamiltonian and then the variation of the spin Hamiltonian, $\delta \mathcal{H}$, can be compared with the variation of the grand canonical potential, $\delta \Omega$. One can make small rotations of the spins at site $i$ and $j$, which is equivalent to a one-site rotation at site $i$ and another one-site rotation at site $j$, while an extra --- interacting --- term also appears due to the fact that the rotations are made simultaneously. This interacting term, which is given by Eq. (5.43) in Ref. \onlinecite{szilva_quantitative_2022}, gives a direct way to derive the exchange formulas. Note that the derivation of the LKAG exchange formula in the seminal paper of Ref.\cite{liechtenstein_local_1987} was also based on the two-site variation strategy. For this reason, we follow here the same strategy for the case of DMI vectors. After the derivation, with the sign convention of Eq.~\ref{ham}, one gets for the DMI term that 
\begin{equation}
	\vec{D}_{ij}=\frac{4}{\pi}\Re\int \mathrm{Tr}_{L }\left(B_i\, G^{00}_{ij}\,B_j\,\vec{G}^{1}_{ji}+B_i\, G^{01}_{ij}\,B_j\,\vec{G}^{0}_{ji}\right)\mathrm{d}\varepsilon \,,
	\label{defD2}
\end{equation}
where $B_{i}$ is introduced by Eq. (5.1) in Ref.~\cite{szilva_quantitative_2022} by utilizing the fact that under small perturbations the responding perturbation in the electronic potential, which is purely spin-dependent, can be divided into local changes of the spin polarised potential in a given region around the atomic sites where the moments are varied. Note that in this paper we mainly focus on the DMI interaction; however, the explicit derived expressions for the Heisenberg $J_{ij}$ and the symmetric anisotropic interaction $\mathcal{A}_{ij}$ can be also found in Ref. \onlinecite{szilva_quantitative_2022}.

\section{Results}
\label{sec3}
\subsection{Magnetic ground states}
In order to investigate and quantify the different contributions to the Heisenberg exchange interactions as well as the DMI, we here present a systematic study of magnetic dimers on several nonmagnetic surfaces. Specifically, we performed density functional theory (DFT) calculations, using the RS-LMTO-ASA method (see Computational details in Appendix \ref{app:C}), for Cr, Mn, and Fe dimers on surfaces where spin-orbit effects are expected to be significant (Pt(001) and W(001)) or weak (Cu(001)).

Initially, we performed self-consistent calculations to determine the electronic and magnetic ground states of the systems studied. Details of this calculation are given in Appendix~\ref{app:C}. All dimers are found to have a canted magnetic configuration as a ground state, either close to AFM or FM, as shown in Table~\ref{tab:config}.

\begin{table}[!htbp]
\centering
\begin{tabular}{
    c
    S[table-format=-2.2]
    S[table-format=1.2]
    S[table-format=3.0, angle-symbol-over-decimal]
}
\toprule
& {$J_{12}$} & {$D_{12}$} & {$\theta_{12}$ (\si{\degree})} \\ 
\midrule[\heavyrulewidth]
Cr$_2$/Cu(001) & -26.56 & 0.03 & 180 \\ \midrule
Mn$_2$/Cu(001) & -3.20 & 1.72 & 174 \\ \midrule
Fe$_2$/Cu(001) & 9.56 & 0.50 & 1 \\ \midrule \midrule
Cr$_2$/Pt(001) & -5.74 & 0.10 & 175 \\ \midrule
Mn$_2$/Pt(001) & -2.31 & 0.14 & 180 \\ \midrule
Fe$_2$/Pt(001) & 2.49 & 1.20 & 7 \\ \midrule \midrule
Cr$_2$/W(001) & 0.88 & 0.13 & 5 \\ \midrule
Mn$_2$/W(001) & 1.31 & 0.15 & 357 \\ \midrule
Fe$_2$/W(001) & -0.07 & 0.10 & 205 \\ 
\bottomrule
\end{tabular}
\caption{The table shows the values of $J_{12}$ and $D_{12}=|\Vec{D}_{12}|$ (mRy) calculated from their respective ground state. Here, $\theta_{12}$ is the angle between the spin moments of atom $i$ and $j$.}
\label{tab:config}
\end{table}

After determining the magnetic ground state for each dimer through our self-consistent process, we calculated the magnetic interactions using the ground state as the reference state. Note that the amplitude of $\theta_{12}$ is proportional to $\frac{|\vec{D}_{12}|}{J_{12}}$. This proportionality is the highest for Fe on W(001), where the total DMI is actually stronger than the isotropic exchange.

\subsection{Relaxation and band-filling effects}
\begin{figure}[htp]
    \centering
    \includegraphics[width=1\linewidth]{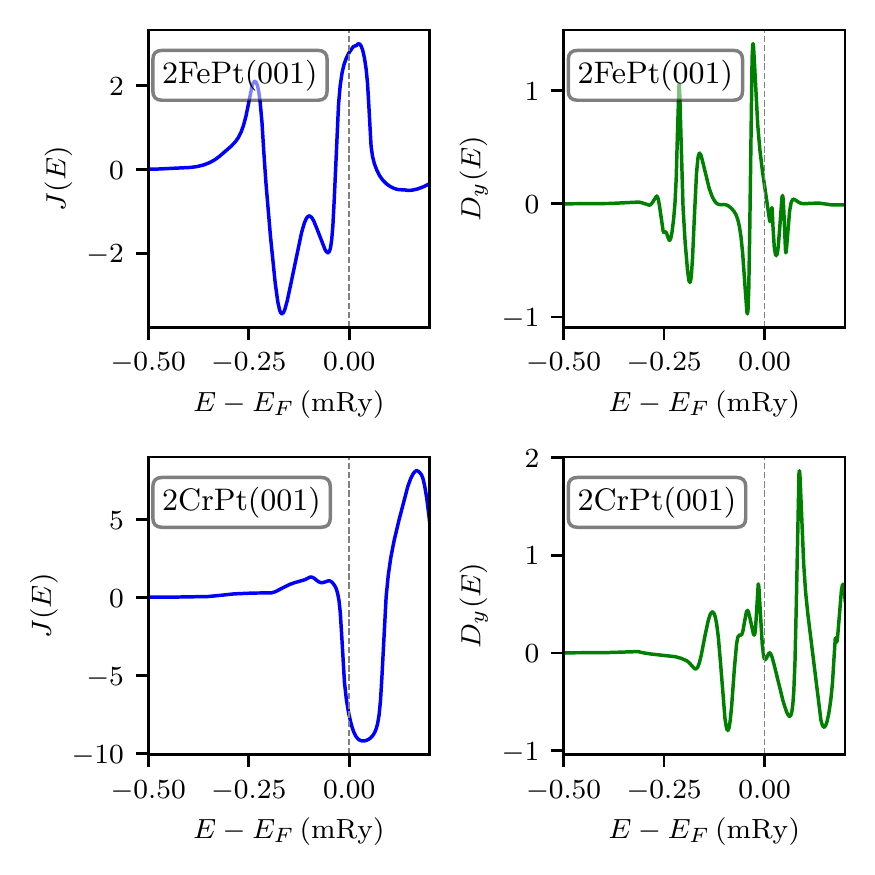}\\
    \caption{Energy dependence of the magnetic interactions $J(E)$ and $D_{y}(E)$. The values for $E=E_{F}$ are the actual values for the calculated magnetic interactions. The interactions are calculated using the ferromagnetic case as a reference.}
    \label{fig:relaxdimers}
\end{figure}

The magnetic interactions of Cr and Fe dimers on Pt(001) have also been also calculated in Ref~\onlinecite{Brinker_2019}. In order to compare their findings with our calculations, structural relaxation must be considered. In that work, the authors obtained that the dimers relax approximately 30$\%$ towards the surface, which can considerably change both $J_{12}$ and $\vec{D}_{12}$. Since a structural relaxation can alter the hybridization and the charge transfer between the deposited dimers and the substrate, one can analyse the magnetic exchange interaction as a function of the band-filling and from there infer the sensitivity of such interactions. An estimate of the band-filling effect can be obtained by looking at the energy dependence of the exchange interactions close to the Fermi Energy.  We have calculated both $J_{12}$ and $\vec{D}_{12}$  as a function of energy for Cr and Fe on Pt(001) and the result is shown in Fig~\ref{fig:relaxdimers}.

The $J_{12}(E)$ and $D_{12}(E)$ curves for both systems, shown in Fig.~\ref{fig:relaxdimers}, exhibit steep slopes around the Fermi Energy, which suggest that the exchange interactions are indeed highly sensitive to relaxation effects. In fact, even a limited change of the band-filling can result in a change of the sign of the Heisenberg exchange of the Fe dimer as well as of the DMI for both dimers, which can effectively alter the magnetic ground state of the dimer. Indeed, when performing calculations for Fe and Cr dimers on a Pt(001) surface with a $\sim$ 35$\%$ inward relaxation, we found that the $J_{12}$ sign varies in comparison to the unrelaxed systems (data not shown). This suggests antiferromagnetic and ferromagnetic couplings, respectively, consistent with the findings in Ref.\onlinecite{Brinker_2019}. The same effect was found for a Mn nanochain on Au(111)~\cite{PhysRevB.93.014438}. Nevertheless, the main aim of this paper is to study the microscopic origin of the DMI interaction for different magnetic configurations and the effects of structural relaxation would not change the conclusions reached in this work. %Hence, we will continue the analysis without considering the relaxation. 

\subsection{Collinear vs noncollinear currents}

%The DMI is known to be present in systems with structural inversion symmetry breaking and spontaneous spin-orbit coupling (SOC) and that the latter induces a intrinsic spin-current into the system~\cite{kikuchi_dzyaloshinskii-moriya_2016} where the DMI can be derived by the linear contribution in SOC of the ground state spin-current~\cite{freimuth_relation_2017}. Another mechanisms responsible to induce spin- and charge-current into the system are noncollinearity and noncoplanarity~\cite{kikuchi_dzyaloshinskii-moriya_2016,tatara_microscopic_2008}, respectively. In the case of the dimer, the coplanarity is always null and, therefore, in this case, only the spin-current adds an extra term to the total DMI in the noncollinear case. While the spin- and charge-current induced by SOC depend on the crystal symmetry, the spin- and charge current induced by noncollinearity depend on the particular choice of rotational plane. This way, by forcing a particular rotational sense, one can analyse these different mechanism separately. We show that while the SOC induced DMI has a weak magnetic configuration dependence, the noncollinear induced DMI has a strong magnetic configuration dependence. Our results show that the configuration dependence of the DMI, both SOC induced and noncollinear induced, is a consequence of the emergence of high-order terms in the spin Hamiltonian, already pointed out in Ref.~\cite{dos_santos_dias_proper_2021}.

\begin{figure}[htp]
\begin{center}
  \includegraphics[width=0.5\linewidth]{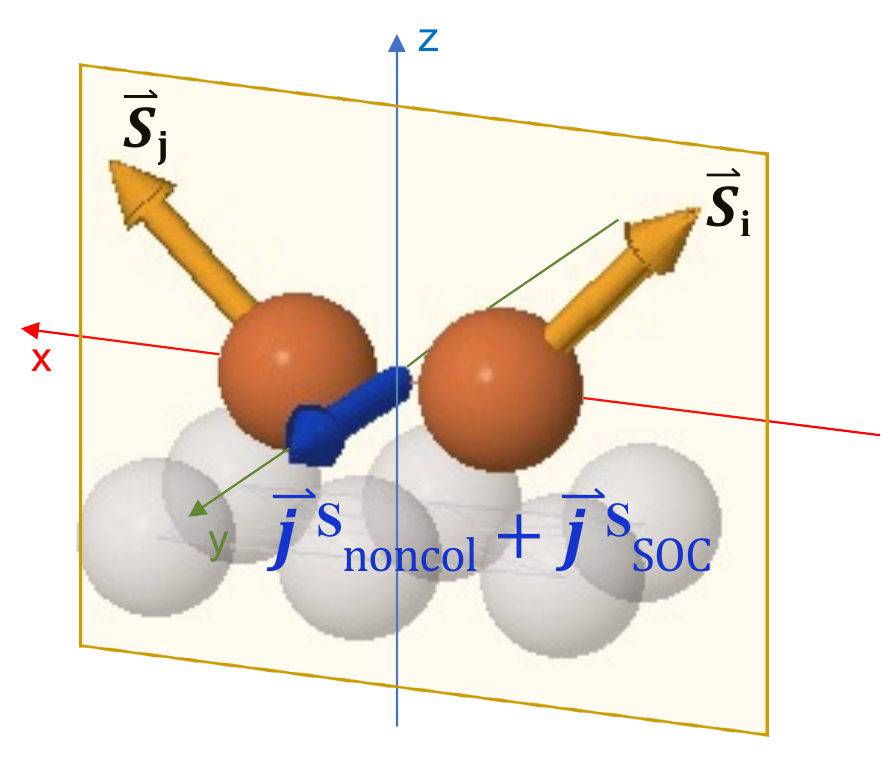}
  %\caption{Magnetic moments rotating in the xz-plane}
  \includegraphics[width=0.5\linewidth]{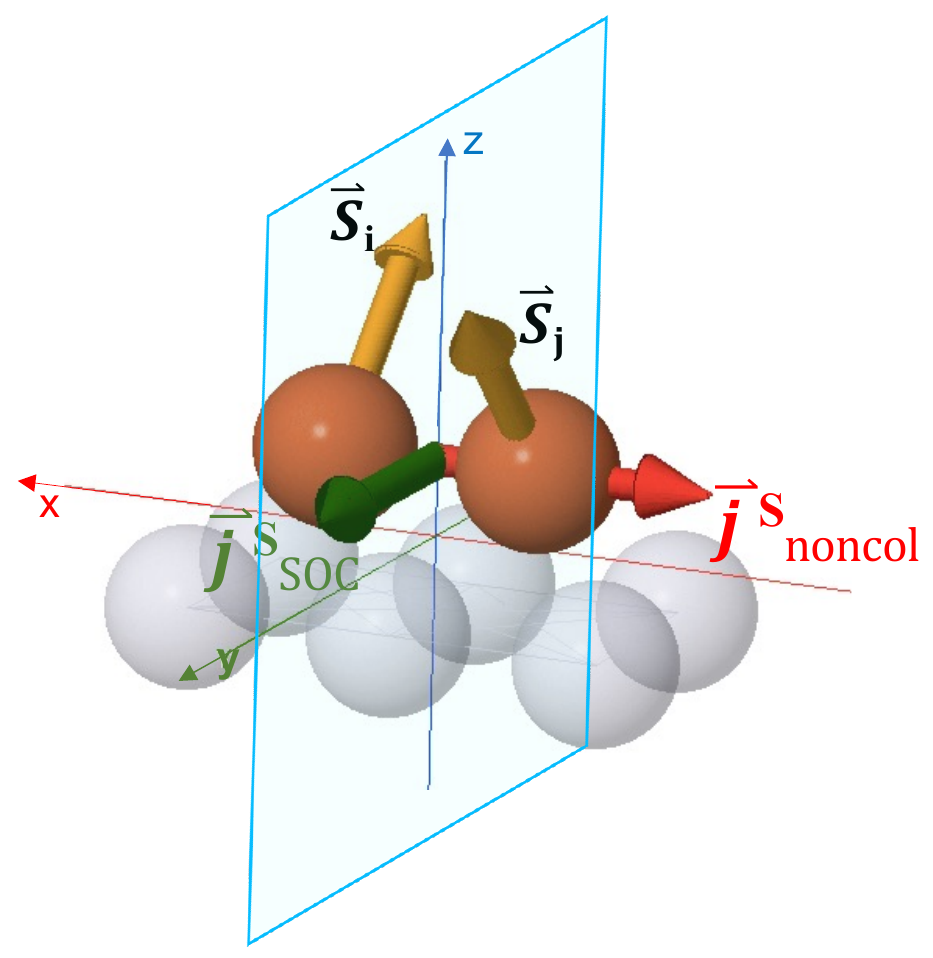}
  %\caption{Magnetic moments rotating in the yz plane}
\end{center}
\caption{
Illustration of the spin-currents in a magnetic dimer. The dimer bond axis is along the x-direction while the z-axis is along the surface normal. The atoms' spin moments ($S_{i}, S_{j}$) are represented by arrows. On top, the plane $xz$ containing the noncollinear magnetic structure is parallel to the bond, and on bottom, the spin moments are rotating in the $yz$ plane, perpendicular to the bond. $\vec{j}_{noncol}$ and $\vec{j}_{SOC}$ denote the spin-currents induced by the noncollinearity of the spin moments and the SOC, respectively.
}
\label{fig:currents}
\end{figure}

It is known that the DMI emerges due to an intrinsic spin-current induced by the SOC~\cite{kikuchi_dzyaloshinskii-moriya_2016}. This can be seen in Eq.~\ref{defD2}, since the spin-current part of the Green function appears explicitly. In this formalism, the DMI is split into two different contributions: $D^{S}\propto G_{ij}^{\eta 1}$, representing the spin-current induced contribution, and $D^{C}\propto G_{ij}^{01}$, representing the charge-current induced contribution, respectively. In the presence of a spin phase difference between the two spin moments, i.e. a noncollinear magnetic configuration, along with electron coherence, leads to a spin current flowing between the atoms. A key aspect of this behaviour is the noncommutativity of the SU(2) spin algebra~\cite{PhysRevB.67.113316}. In the dimer case, this leads to a spin-current polarized perpendicularly to the plane of rotation. This spin-current is spontaneous and induces a torque on the spin moments analogously to what the spin current induced by SOC does. These torques play a significant role in determining the system's magnetic ground state. In case of a triangular trimer, the noncollinearity also gives rise to a charge current, if the scalar spin chirality $(\vec{e}_{i}\times\vec{e}_{j})\cdot\vec{e}_{k}$ is different than zero~\cite{PhysRevB.67.113316}. However, since our study focuses solely on dimers, the $D^{C}$ has a minimal impact on the total DMI. This is explored in more detail in Appendix \ref{app:A}, in Eq.~\ref{eq-simDMI} where we obtain that
\begin{equation}
    \vec{j}_{s} \propto (\vec{e}_{1}\times\vec{e}_{2}).
    \label{equation:jsnoncol}
\end{equation}
This spin-current induced by the noncollinearity introduces an extra term in the DMI, leading to two distinct contributions. For example, we considered two rotations in different planes, depicted in Fig.~\ref{fig:currents}: (top) in the xz plane and in (bottom) yz plane. For a (001) surface, if the bond between the two atoms is the x-axis, a DMI is found in the y-direction $D_{y}$ according to Moriya rules~\cite{crepieux_dzyaloshinskymoriya_1998}. If the rotation described in Fig.~\ref{fig:currents}(a) is performed, a spin current in the y-direction emerges and then one can see a total DMI with both contributions, the SOC and the noncollinear, being in the y-direction. Conversely, if the rotation shown in  Fig.~\ref{fig:currents}(b) is done instead, the spin-current induced by the noncollinearity, according to Eq.~\ref{equation:jsnoncol}, is in the x-direction and then one can see two independent contributions that are not parallel. For the rotation of the spin moments specified in case (b), one can analyse the contributions separately.

In Appendix \ref{app:B} we derive the reference dependence of both the two independent types of Dzyaloshinksii-Moriya interactions, that differ in the origin of the spin currents, i.e.,~either noncollinearity or spin-orbit coupling.

%Because, for this example, the $D_{y}$ has no dependence on $\xi^{0}$, with $\xi$ being the SOC strength, it is immediately 0 in absence of SOC for any given magnetic configuration. In other hand, $D_{x}$ will have dependence of $\xi^{0}$ and higher-orders, which makes it to not vanish in the absence of SOC, but only present in the case of noncollinearity. In this way, one can designate $D_{y}$ as the usual DMI, being entirely dependent on SOC and zero if SOC is not included in the calculation; and $D_{x}$ as the part of the DMI that is strongly induced by noncollinearity, but also by SOC.
%%%%%%%%%%%%%%
%The Fig.~\ref{dimer-scale} refers exactly to that case (b). Firstly, one can see that both contributions scale differently with SOC. While $D_{y}$ scales linearly, $D_{x}$ scales quadratic. The $D_{y}$ is the usual DMI being fully dependent on SOC and zero if SOC is turned off (dashed line). The $D_{x}$ is the part of the DMI that is strongly induced by noncollinearity but also by SOC, as it can be seen that the dashed and full lines do not fall perfectly on top of each other.

\subsection{Dzyaloshinskii-Moriya interactions}

\begin{figure*}[htp]
    \begin{center}
    \includegraphics[width=1\linewidth]{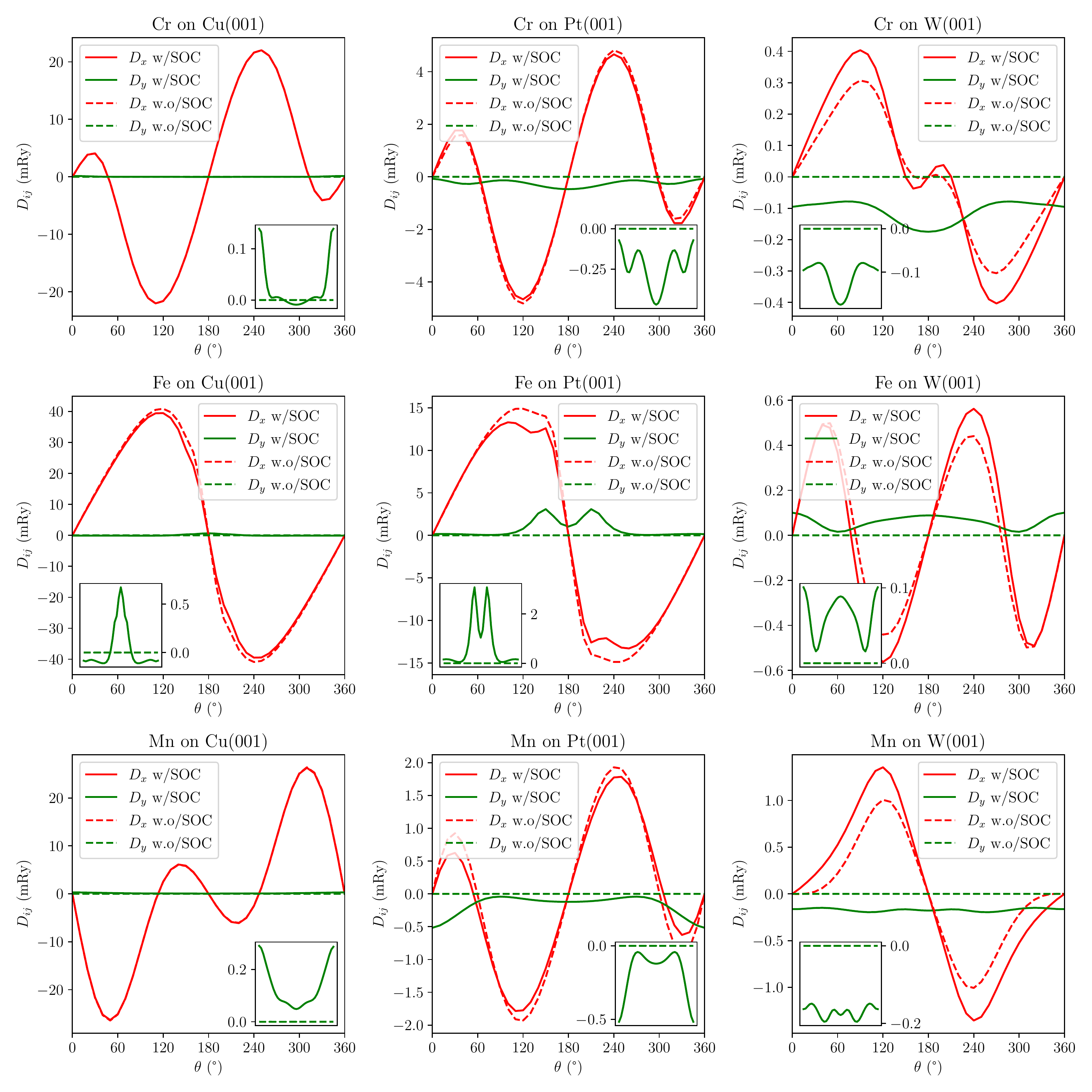}
    \end{center}
    \caption{DMI calculated when varying the angle $\theta$ between the spin moments of the dimer atoms around the dimer bond axis (x-direction). The top panels display the results for Cr dimers on Cu(001), Pt(001), and W(001) from left to right, while the middle panels refer to the Fe dimers. The bottom panels refer to the Mn dimers on the same substrates. In each panel, the red lines denote the DMI component in the bonding direction of the atoms, $D_{x}$, while the green lines denote the DMI component perpendicular to the bond, $D_{y}$. The full lines stand for calculations when the spin-orbit coupling is included, whereas the dashed lines denote calculation without spin-orbit coupling (SOC). The insets represent a zoom into the $D_{y}$ component.}
    \label{3dwptcu}
\end{figure*}

%%%
 In order to understand the dependence of the DMI with respect to both the angle between the spin moments of the dimer atoms, $\theta$, as well as to the SOC, we performed calculations of the DMI considering the rotational plane shown in Fig.~\ref{fig:currents}(bottom). In this configuration, the different contributions to the DMI are perpendicularly aligned and can be studied separately. We consider one spin moment fixed along the z-axis while the other one rotates around the x-axis with an angle $\theta$. In this case, the noncollinearity will induce a spin-current in the x-axis giving rise to what we have termed a DMI-like contribution to the exchange interaction~\cite{cardias_dzyaloshinskii-moriya_2020,cardias_first-principles_2020}. A finite value of the SOC will induce a spin-current parallel to the mirror plane between the atoms giving a finite DMI along the y-axis. The calculated angular dependences of $D_{x}$ and $D_{y}$ are shown in Fig.~\ref{3dwptcu}. One can see that in the absence of SOC (dashed line), the $D_{y}$ is immediately zero, while the DMI-like term ($D_{x}$) is still finite. Taking Eq.~\ref{ham}, we can write the energy coming from each term as
\begin{align}
    E_{D_{x}} =  D_{x}(\vec{e}_i\times\vec{e}_{j})^{x} \label{eq:dx} \\ 
    E_{D_{y}} =  D_{y}(\vec{e}_i\times\vec{e}_{j})^{y},
    \label{eq:dy}
\end{align}
where the upper script $x$ and $y$ designates the corresponding cartesian component of the vector product between the two spin moments. It is possible to see in Fig.~\ref{3dwptcu} that the $D_{y}$ has the property of $D_{y}(\theta)=D_{y}(-\theta)$, while the DMI-like term $D_{x}$ does not, i.e. $D_{x}(\theta)=-D_{x}(-\theta)$. According to Eqs~\ref{eq:dx} and \ref{eq:dy} combined with the fact that the vector product is odd in $\theta$, $E_{D_{x}}$ will have the same value for both $\theta$ and $-\theta$, 
%since $\vec{e}_i\times\vec{e}_{j}=\hat{y}\sin\theta$ is odd in $\theta$, 
while $E_{D_{y}}$ will have the same magnitude but a different sign. It means that in this particular setup, $D_{y}$ is responsible to lift the degeneracy between two different rotational senses (chiral interaction), while $D_{x}$ acts as an extra term to the total energy coming from the $\vec{e}_{i}\times\vec{e}_{j}$ contribution. It is important to point out that both the DMI ($D_{y}$) and the DMI-like term ($D_{x}$) are configuration-dependent, but only $D_{y}$ is fully dependent on SOC. The substrates Cu, Pt, and W have an ascending strength of SOC, which does not necessarily explicitly translate to an ascending magnitude of the $D_{y}$. This is expected since the DMI is a result of a complex interplay of various factors beyond the strength of SOC, e.g. band filling~\cite{belabbes_hunds_2016}. On the other hand, the DMI-like $D_{x}$ term is seen to have an inversely proportional relationship with the SOC, being the weakest for the dimers on W(001). In the particular case of the dimer, the spin-current that flows between $\vec{e}_{i}$ and $\vec{e}_{j}$ induces a spin accumulation proportional to $\vec{e}_{i}\times\vec{e}_{j}$, which induces a torque into the spin moments contributing to the final magnetic ground state. Note that, in general, the DMI ($D_{y}$) is only weakly dependent on the magnetic configuration, which should reflect in just a small variance of the ground state angle between the spin moments. However, in some special cases e.g. Cr and Fe on Cu(001), the DMI changes sign. In this case, it means that the system has a different chirality which is analogous to the change of FM to AFM ordering if the isotropic exchange goes from positive to negative, respectively, as seen in the Fe dimer case~\cite{streib_adiabatic_2022}.

\subsection{From local to global representation}

\begin{figure*}[htp]
    \centering
    \includegraphics[width=1\linewidth]{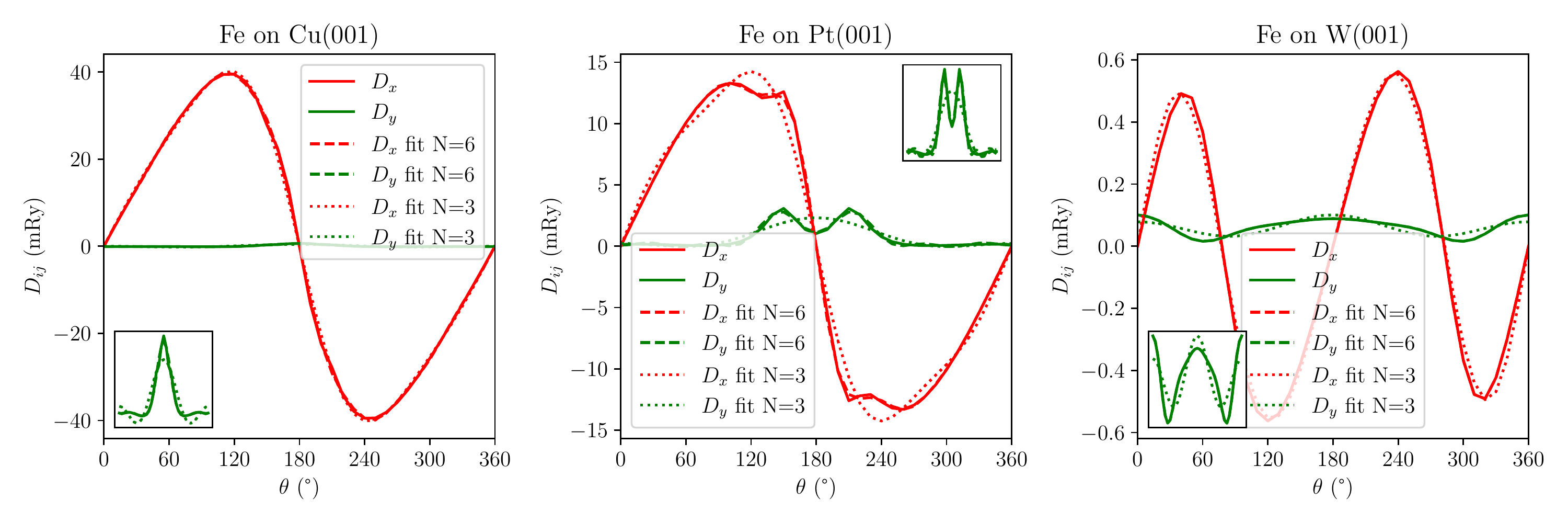}\\
    \caption{
%DMI fit, 3d dimers on W(001), Pt(001) and Cu(001). Improve description.
Fitting the self-consistent calculations (configuration-dependent magnetic interactions) with the multi-spin representation (Eqs.~\ref{eq:multispin1} and \ref{eq:multispin2}. Here, the full lines are the DFT calculations for the DMI when the spin-orbit coupling is included, while the dashed and dotted curves represent the fitted results of the DFT calculations for N=3 and 6, respectively, using Eqs. 3.2 and 3.3. Note that in the presented examples, the full line and the dashed line are almost indistinguishable while the dotted lines are slightly off and not a good fit in some cases. 
}
    \label{fig:fit3dwptcu}
\end{figure*}

The spin Hamiltonian, commonly known as the Heisenberg Hamiltonian is described by a set of parameters which are expected to be sufficient to describe the whole energy landscape of a given system, such as that the total energy is a function of the spin moment directions only, $\vec{e}_{i}$, as described in Eq.~\ref{ham}. There are several works in the literature where a simple bilinear spin Hamiltonian is not enough to describe the total energy, therefore, high-order terms can be needed as a correction to the Hamiltonian. For instance, if only the isotropic exchange is considered for a simple dimer, the total energy should be described as a linear function of $\cos(\theta)$ i.e. as $E_{\text{dimer}}=-J\cos(\theta)$, where $J$ is the isotropic exchange and $\theta$ is the angle between the spin moments of the atoms. 
%{\bf Ramon to be consistent with Eqn.1.1 I suggest we skip the minus sign here and in the expression below. Ramon: fixed}
In certain scenarios, as shown in Ref.~\cite{PhysRevLett.94.137203}, higher order terms such as $\cos^{2}(\theta)$ are important. Consequently, the energy is better described as $E_{\text{dimer}}=-J_{1}\cos(\theta) - J_{2}\cos^{2}(\theta)$, where $J_{2}$ is now a biquadratic term. This was also the case found for an Fe dimer~\cite{cardias_spin_2021,streib_adiabatic_2022}, a chiral biquadratic pair interaction~\cite{Brinker_2019} and more complex systems such an Fe monolayer on Ir(111)~\cite{heinze_spontaneous_2011}. Therefore, to find the correct Hamiltonian can be a complicated task. 

The concept of a local Hamiltonian is discussed in Refs.~\cite{streib_exchange_2021,szilva_quantitative_2022,brinker_prospecting_2020} and relies on the magnetic parameters calculated from a given magnetic reference state. The magnetic interactions are, in this description, a function of the magnetic texture of the system. The positive aspect of this approach is that only a bilinear Hamiltonian is needed. In Appendix~\ref{app:B}, we demonstrate how the local and global descriptions relate to each other. For the purpose of the present work, we write explicitly how the reference state dependent bilinear DMI can be re-expressed as a combination of multi-spin interactions, which in the current case of dimers simplifies to higher order interactions of the two spin moments as

\begin{align}
    D_{\text{noncol}} &= \sin\theta\sum_{n=0}^N{a_{n}\cos^{n}\theta} \label{eq:multispin1}
    \\
    D_{\text{SOC}} &= \sum_{n=0}^N{b_{n}\cos^{n}\theta},
    \label{eq:multispin2}
\end{align}
where $\theta$ is the angle between the two spin moments of the sites being considered, $a_{n}$ and $b_{n}$ are %the multi-spin 
parameters that capture the multi-spin character of the reference dependence. The nonrelativistic $a_n$ parameters arise from the reference dependence of the spin current which in turn originates from the vector chirality between the two moments, $\vec{e}_1\times\vec{e}_2=\sin\theta\,\vec{e}_3$, where the unit vector $\vec{e}_3$ is orthogonal to both spin moments,  combined with higher order factors of $\vec{e}_1\cdot\vec{e}_2=\cos\theta$. For the relativistic terms, the $b_0$ parameter corresponds to the usual bilinear case, while those with $n>0$ are again higher order corrections that catch the reference dependence, e.g.~a biquadratic term for $b_1$ and so on. These expressions clearly demonstrate that  $D_{\text{noncol}}$ vanishes if the spin moments are aligned in a collinear ($\theta=0$) fashion  but can be finite if the two spin moments are noncollinear.
Nb.~the factors $a_n$ and $b_n$ are in principle possible to calculate as higher order interactions, but already in the dimer case, this is a cumbersome task, which we leave to future studies.

%Manuel dos Santos Dias \textit{et al.} in
 The authors of Ref.~\onlinecite{dos_santos_dias_proper_2021} hinted that higher order four-spin isotropic interaction of the form $(\vec{e}_{i}\cdot\vec{e}_{j})(\vec{e}_{i}\cdot\vec{e}_{j})$ (adapted for the dimer case) can take form of a DMI-like interaction, achiral, due to a manipulation of the unitary vectors via vector identities operations, which is exactly the case for Eq.~\ref{eq:multispin1} with N=0. However, the microscopic origins of neither interactions were discussed, which we explicitly demonstrate in this work. For instance, the intrinsic spin-current driven by the spin-orbit coupling, $D_{y}$, leads to a chiral interaction while the spin-current induced by noncollinearity, $D_{x}$, leads to an achiral interaction.

We have used two different cutoffs in the expansions, $N=3$ and $N=6$, in Eqs.\ref{eq:multispin1} and \ref{eq:multispin2} when fitting the results of our DFT calculations. The results are shown in Fig.~\ref{fig:fit3dwptcu}. In order to improve readability, we have chosen only the cases where the differences between the fits for different $N$'s are more significant, which are the cases of Fe on Cu(001), Pt(001), and W(001). The dotted and dashed curves represent the fit where optimal values for the parameters are obtained so that the sum of the squared residuals is minimized, while the full lines are our DFT calculations. While results with $N=6$ fit well with all examples considered, the fits with $N=3$ are not fully converged for several cases. It suggests that the determination of a proper spin-Hamiltonian is highly dependent on the system considered.

\section{Conclusion}

To summarize, in the present work we have presented a detailed analysis of the microscopic origin of the Dzyaloshinskii-Moriya (DM) and the DM-like exchange interaction that arises from non-collinear magnetic configurations. We show that while both are influenced by the spin-orbit coupling (SOC), the former is only finite under the presence of SOC and the latter is finite even without SOC, but requires instead non-collinear magnetism. For that reason it has been named non-collinear DMI interaction \cite{cardias_first-principles_2020,cardias_comment_2022}. To quantify the analysis presented here, we calculated the electronic structure of dimers (Cr, Fe, and Mn) on Pt(001), W(001), and Cu(001) surfaces using the first principles RS-LMTO-ASA method. Based on these calculations, we computed the DMI for various magnetic configurations. Firstly, by employing a formalism that enables the separation of the DMI contributions distinguishing spin- and charge current induced DMI terms, we have clarified that the usual spin-orbit driven DMI component is induced by the spin current generated by the spin-orbit coupling, while the non-collinear DM-interaction is induced by the non-collinearity of the spin moments. Given the absence of the chiral behavior of the latter interaction, it is clear that it gives no contribution to a preferential rotational sense, unlike the conventional spin-orbit induced DMI.

We also addressed the interpretation of a spin Hamiltonian and the connection between magnetic configuration dependent interactions and a multi-spin approach, where we argue that both are complementary. By doing so, we explicitly show that the dependence of magnetic configuration on these interactions can be mapped onto multi-spin parameters that are independent (or at least less dependent) of the underlying magnetic configuration. The caveat of using this approach is that the Hamiltonian needed, and therefore the complete set of parameters, is not defined a priori. Also, from such an analysis, it follows that one needs in principle to update (recalculate) the magnetic parameters at every time-step to be used e.g. in spin-dynamics simulations, if one uses a simpler (e.g. bilinear) spin Hamiltonians. In order to avoid this constant update, one could use a more complex spin Hamiltonian with all the relevant multi-spin interactions (biquadratic etc.) included. To calculate the latter interactions can be tricky, although bilinear and biquadratic interactions are dominant in most of the systems reported in the literature. Most likely, systems that present a complex magnetic texture as a ground state, such as skyrmions and other non-collinear antiferromagnet systems (as can be found in Mn$_{3}$X, X=Sn, Ge, Ge and  Mn$_{3}$Y, Y= Pt, Ir or Rh) are strong candidates to have finite and important multi-spin interactions.

The findings presented here also shed light on the underlying mechanisms of the Dzyaloshinskii-Moriya interactions. By explaining the relation between the geometry of the spin moment orientation and the emergence of new interactions, and its microscopic origin, we hope to provide insights into the design and control of magnetic materials for spintronics applications. For instance, it might help with new strategies when using effects that can either generate or inject spin-currents into the system, such as the Spin Seebeck effect or Edelstein effect, that can be used to transfer spin-current through surfaces when tailoring new magnetic materials.
 
\section{Acknowledgements}
R.C acknowledge financial support from FAPERJ - Fundação Carlos Chagas Filho de Amparo à Pesquisa do Estado do Rio de Janeiro, grant number E-26/205.956/2022 and 205.957/2022 (282056). A.B.K. and J.S.S. acknowledge financial support from CAPES and CNPq Brazil. 
%%%%%%%%
A.B.K. acknowledges the INCT of Materials Informatics.
%%%%%%%
Valuable discussions with Dr. Danny Thonig and Prof. Mikhail Katsnelson are acknowledged. O.E. acknowledges support from the Swedish Research Council (VR), the European Research Council (ERC, FASTCORR project), the Knut and Alice Wallenberg Foundation (KAW) and STandUPP.
O. E. and A. D. acknowledge support from the Wallenberg Initiative Materials Science for Sustainability (WISE) funded by the Knut and Alice Wallenberg Foundation (KAW). O.E. and A. B. acknowledge financial support from eSSENCE. 
A.D. acknowledges financial support from
the Swedish Research Council (Vetenskapsrådet, VR) Grant Nos. 2016-05980 and 2019-05304, and the Knut and Alice Wallenberg (KAW) Foundation Grant Nos. 2018.0060, 2021.0246, and 2022.0108.
L.N. acknowledges support from VR.
The computations were enabled by resources provided by the computational
facilities of the CCAD/UFPA (Brazil), at the National Laboratory for Scientific Computing (LNCC/MCTI, Brazil) as well as the National Academic Infrastructure for Supercomputing in Sweden (NAISS) and the Swedish National Infrastructure for Computing (SNIC) at NSC and PDC, partially funded by the Swedish Research Council through grant agreements no. 2022-06725 and no. 2018-05973.
J.F. acknowledges support from Vetenskapsr\aa det and Stiftelsen Olle Engkvist Byggm\"astare.

\appendix

\section{Analytical dimer model}
\label{app:A}
The purpose of considering an analytical model is to demonstrate in a clear way that magnetic impurities in an otherwise nonmagnetic electronic structure, give rise to the anti-symmetric anisotropy which is the scope of the present article. Here, we may controllably change the premises of an underlying two-dimensional electron gas by tuning the intrinsic spin-orbit coupling. In particular, we demonstrate that an intrinsic spin-orbit coupling is not required for the existence of an anti-symmetric magnetic anisotropy between two magnetic impurities embedded in the electron gas.
\subsection{Defects on a metallic surface}
We assume a simple two-dimensional electron gas with a Rashba spin-orbit coupling, e.g., surface states on a metallic surface, in which magnetic defects are embedded. We model this system by the Hamiltonian
\begin{align}
\Hamil=&
	\sum_\bfk\Psi^\dagger_\bfk\bfepsilon_\bfk\Psi_\bfk
	+
	\int\Psi^\dagger(\bfr)\bfV(\bfr)\Psi(\bfr)d\bfr,
 \label{eq-Ham}
\end{align}
where the spinor $\Psi_\bfk=(\cs{\bfk\up}\ \cs{\bfk\down})^t$ ($\Psi^{\dagger}_\bfk=(\csdagger{\bfk\up}\ \csdagger{\bfk\down})$) denotes the annihilation (creation) operator for electrons with crystal momentum $\bfk$ and spin $\sigma=\up,\down$, defined by the energy spectrum captured in the matrix $\bfepsilon_\bfk=\dote{\bfk}\sigma^0+\alpha[\bfk\times\hat{\bf z}]\cdot\bfsigma$ at the momentum $\bfk$, where $\alpha$ denotes the Rashba spin-orbit coupling strength. Moreover, the $\Psi(\bfr)=\int\Psi_\bfk e^{-i\bfk\cdot\bfr}d\bfk/\Omega$ and $\Psi^\dagger(\bfr)=\int\Psi^\dagger_\bfk e^{i\bfk\cdot\bfr}d\bfk/\Omega$ are the corresponding coordinate space operators, where $\Omega$ is the integration volume, whereas the scattering potential $\bfV(\bfr)=\sum_m\bfV_m\delta(\bfr-\bfr_m)$ defines a collection of defects $\bfV_m=V_m\sigma^0+\bfm_m\cdot\bfsigma$.

\subsection{Green function}

In this model, the unperturbed retarded Green function (GF) $\bfg_\bfk$ is defined for the first term of Eq. \eqref{eq-Ham}, and is given in reciprocal and real space by the expressions
\begin{subequations}
\begin{align}
\bfg_\bfk(z)=&
	\frac{(z-\dote{\bfk})\sigma^0+\alpha[\bfk\times\hat{\bf z}]\cdot\bfsigma}{(z-\dote{\bfk})^2-\alpha^2k^2}
    ,
\\
\bfg(\bfr,\omega)=&
	-i\frac{N_0}{4}
	\sum_{s=\pm}
	\frac{\kappa_s}{\kappa}
	\Bigl(
		H_0^{(1)}(\kappa_sr)\sigma^0
\nonumber\\&
		-
		isH_1^{(1)}(\kappa_sr)[\hat\bfr\times\hat{\bf z}]\cdot\bfsigma
	\Bigr)
    ,
\end{align}
\end{subequations}
respectively. These two representations of $\bfg$ are related via the Fourier transform $\bfg_\bfk=\int\bfg(\bfr)e^{i\bfk\cdot\bfr}d\bfr$. Moreover, we have introduced the notation $\kappa=\sqrt{2N_0(\omega+\dote{F}-\alpha^2N_0/2)}$, where $\dote{F}$ denotes the Fermi energy, $\kappa_s=\kappa+s\alpha N_0$, and $N_0=m/\hbar^2$, whereas $H_m^{(1)}$ is a Hankel function. Also, $k=|\bfk|$. In this way we have defined $\bfg_\bfk=g_0\sigma^0+\bfg_1\cdot\bfsigma$ in both coordinate and reciprocal space.

The dressed GF $\bfG$ is calculated by the inclusion of the scattering potential $\bfV$, c.f., Eq. \eqref{eq-Ham}. The dressed GF can be formulated in terms of the $T$-matrix expansion of the impurity potential, giving in reciprocal space the expression
\begin{align}
\bfG_{\bfk\bfk'}=&
	\delta(\bfk-\bfk')\bfg_\bfk
	+
	\sum_{mn}
		\bfg_\bfk e^{-i\bfk\cdot\bfr_m}\bfT(\bfR_{mn})e^{i\bfk'\cdot\bfr_n}\bfg_{\bfk'},
\end{align}
where $\bfR_{mn}=\bfr_m-\bfr_n$, whereas the $T$-matrix is given by
\begin{subequations}
\begin{align}
\bfT(\bfR_{mn})=&
	\bfV_m(\bft^{-1})_{mn}
	,
\\
\bft_{mn}=&
	\delta_{mn}\sigma^0-\bfg(\bfR_{mn})\bfV_n
	.
\end{align}
\end{subequations}
For later purpose, we also define the correction $\delta\bfG(\bfr,\bfr')$ to the Green function in coordinate space (via $\bfG(\bfr,\bfr')=\int\bfG_{\bfk\bfk'}e^{-i\bfk\cdot\bfr+i\bfk'\cdot\bfr'}d\bfk d\bfk'/\Omega^2$) as
\begin{align}
\delta\bfG(\bfr,\bfr')=&
	\sum_{mn}
		\bfg(\bfr-\bfr_m)\bfT(\bfR_{mn})\bfg(\bfr_n-\bfr')
	.
\end{align}
Here, since the scattering potential is partitioned into a nonmagnetic and a magnetic component, we can write $\bfT(\bfR_{mn})=T_0(\bfR_{mn})\sigma^0+\bfT_1(\bfR_{mn})\cdot\bfsigma$. We also remark that the potential $\bfV_m$ should here be considered as an expectation value, such that $\bfV_m=V_m\sigma^0+\av{\bfm_m}\cdot\bfsigma$.

Next, we show that the $T$-matrix is not necessarily symmetric under changes of the site order $\bfR_{mn}\rightarrow\bfR_{nm}$. For the sake of argument, consider two magnetic impurities on the surface, giving
\begin{align}
\bft=&
	\begin{pmatrix}
		\sigma^0-g_0\bfV_1 & -\bfg(R_{12})\bfV_2 \\
		-\bfg(R_{21})\bfV_1 & \sigma^0-g_0\bfV_2
	\end{pmatrix}
	,
\end{align}
where each entry is a $2\times2$-matrix. The diagonal components only include $g_0$ since $\bfg_1(R_{mm})\equiv0$. Moreover, we denote $R_{mn}=|\bfR_{mn}|$, and $g_0=g_0(R_{mm})$, whereas we set $g_{mn}=g_0(R_{mn})$, and $\bfg_{mn}=\bfg_1(R_{mn})$, for $m\neq n$.

We need the inverse of the matrix $\bft$ and, for the sake of notation, set $\bfs=\bft^{-1}$. Then, we can write
\begin{subequations}
\begin{align}
\bfs_1=&
	\biggl(
		\sigma^0-g_0\bfV_1-\bfg(R_{12})\bfV_2\Bigl(\sigma^0-g_0\bfV_1\Bigr)^{-1}\bfg(R_{21})\bfV_1
	\biggr)^{-1}
	,
\\
\bfs_{12}=&
	-\bfs_1\bfg(R_{12})\bfV_2\Bigl(\sigma_0-g_0\bfV_2\Bigr)^{-1}
	,
\\
\bfs_{21}=&
	-\bfs_2\bfg(R_{21})\bfV_1\Bigl(\sigma_0-g_0\bfV_1\Bigr)^{-1}
	,
\\
\bfs_2=&
	\biggl(
		\sigma^0-g_0\bfV_2-\bfg(R_{21})\bfV_1\Bigl(\sigma^0-g_0\bfV_2\Bigr)^{-1}\bfg(R_{12})\bfV_2
	\biggr)^{-1}
	.
\end{align}
\end{subequations}
These expressions can be simplified, without losing any essential feature pertaining to the symmetries of the $T$-matrix, by assuming that
\begin{align}
\biggl\|
	\bfg(R_{12})\bfV_2\Bigl(\sigma^0-g_0\bfV_1\Bigr)^{-1}\bfg(R_{21})\bfV_1
	\biggr\|
	\ll&
	\biggl\|
		\sigma^0-g_0\bfV_1
	\biggr\|
	.
\end{align}
This assumption can be achieved by locating the impurities at a sufficiently large distance from one another, the so-called dilute limit. We also assume that the impurities constitute a purely magnetic scattering potential, hence, $V_m=0$. This simplification may be somewhat unphysical, since any scattering potential would generally comprise the nonmagnetic component $V_m$. However, this simplification is introduced in order to reduce the complexity of the expressions since the presence of $V_m$ does not alter the target expression for the anti-symmetric anisotropy. Then, $\bfs$ reduces to
\begin{subequations}
\begin{align}
\bfs_1=&
	\biggl(
		\sigma^0-g_0\av{\bfm_1}\cdot\bfsigma
	\biggr)^{-1}
	=
	\frac{\sigma^0+g_0\av{\bfm_1}\cdot\bfsigma}{1-g_0^2|\bfm_1|^2}
	,
\\
\bfs_{12}=&
	-
	\bfg(R_{12})
		\frac{\sigma^0+g_0\av{\bfm_1}\cdot\bfsigma}{1-g_0^2|\bfm_1|^2}
		\av{\bfm_2}\cdot\bfsigma
		\frac{\sigma^0+g_0\av{\bfm_2}\cdot\bfsigma}{1-g_0^2|\bfm_2|^2}
	,
\\
\bfs_{21}=&
	-
	\bfg(R_{21})
		\frac{\sigma^0+g_0\av{\bfm_2}\cdot\bfsigma}{1-g_0^2|\bfm_2|^2}
		\av{\bfm_1}\cdot\bfsigma
		\frac{\sigma^0+g_0\av{\bfm_1}\cdot\bfsigma}{1-g_0^2|\bfm_1|^2}
	,
\\
\bfs_2=&
	\frac{\sigma^0+g_0\av{\bfm_2}\cdot\bfsigma}{1-g_0^2|\bfm_2|^2}
	.
\end{align}
\end{subequations}
It is clear that $\bfs_{12}\neq\bfs_{21}$ and, hence, $\bfT(\bfR_{mn})\neq\bfT(\bfR_{nm})$ whenever $\av{\bfm_1}\cdot\bfsigma\neq\av{\bfm_2}\cdot\bfsigma$. For a simpler notation, below we shall write $\bfm_m$ instead of $\av{\bfm_m}$.

The critical components of the off-diagonal $T$-matrix elements are the numerators since the denominators are equal in the two elements. In the given notation, one derives for $m\neq n$
\begin{widetext}
\begin{align}
\bfT(\bfR_{mn})\sim&
	\bfm_m\cdot\bfsigma(\sigma^0+g_0\bfm_m\cdot\bfsigma)\bfg(R_{mn})\bfm_n\cdot\bfsigma(\sigma^0+g_0\bfm_n\cdot\bfsigma)
\nonumber\\=&
	\biggl\{
		g_{mn}\Bigl(g_0^2m_m^2m_n^2+\bfm_m\cdot\bfm_n\Bigr)
%\nonumber\\&
		+
		\bfg_{mn}\cdot
		\Bigl[
			g_0\Bigl(\bfm_mm_n^2+m_m^2\bfm_n\Bigr)
			-
			i\bfm_m\times\bfm_n
		\Bigr]
	\biggr\}
	\sigma^0
\nonumber\\&
	+
	\biggl\{
		g_0^2m_m^2m_n^2\bfg_{mn}
		+
		g_0g_{mn}\Bigl(m_n^2\bfm_m+m_m^2\bfm_n\Bigr)
%\nonumber\\&
		+\bfm_m\cdot\bfg_{mn}\bfm_n
		+
		\bfm_m\bfg_{mn}\cdot\bfm_n
		-
		\bfm_m\cdot\bfm_n\bfg_{mn}
\nonumber\\&
		-
		ig_0\bfg_{mn}\times
		\Bigl(
			\bfm_mm_n^2
			-
			m_m^2\bfm_n
		\Bigr)
		+
		ig_{mn}\bfm_m\times\bfm_n
	\biggr\}
	\cdot
	\bfsigma
	.
\label{eq-2by2Tmn}
\end{align}
%\end{widetext}
Using these explicit expressions for the $T$-matrix, the charge and spin density components $G_{00}$ and $\bfG_{10}$, and the corresponding current components $G_{01}$ and $\bfG_{11}$ can be identified from the charge and spin components $G_0$ and $\bfG_1$, respectively. The two-index notation refers to the charge and spin densities $G_{00} $ and $\bfG_{10}$, and the charge and spin currents $G_{10}$ and $\bfG_{11}$.

In doing so, first identify the components in the $T$-matrix, Eq. \eqref{eq-2by2Tmn}, that are even and odd under $\bfR_{mn}\rightarrow\bfR_{nm}$, using the properties $g_0(-\bfr)=g_0(\bfr)$ while $\bfg_1(-\bfr)=-\bfg_1(\bfr)$. Hence, adopting the two-index notation, the four components to the $T$-matrix are summarized as
%\begin{widetext}
\begin{subequations}
\begin{align}
T_{00}(\bfR_{mn})=&
	g_{mn}\Bigl(g_0^2m_m^2m_n^2+\bfm_m\cdot\bfm_n\Bigr)
%\nonumber\\&
	-i\bfg_{mn}\cdot\Bigl(\bfm_m\times\bfm_n\Bigr)
	,
\\
T_{01}(\bfR_{mn})=&
	g_0
	\bfg_{mn}
	\cdot
	\Bigl(\bfm_mm_n^2+m_m^2\bfm_n\Bigr)
	,
\\
\bfT_{10}(\bfR_{mn})=&
	g_0g_{mn}\Bigl(m_n^2\bfm_m+m_m^2\bfm_n\Bigr)
%\nonumber\\&
	-
	ig_0\bfg_{mn}\times
	\Bigl(
		\bfm_mm_n^2
		-
		m_m^2\bfm_n
	\Bigr)
	,
\\
\bfT_{11}(\bfR_{mn})=&
	g_0^2m_m^2m_n^2\bfg_{mn}
	+
	ig_{mn}\bfm_m\times\bfm_n
	+
	\bfm_m\cdot\bfg_{mn}\bfm_n
%\nonumber\\&
	+
	\bfm_m\bfg_{mn}\cdot\bfm_n
	-
	\bfm_m\cdot\bfm_n\bfg_{mn}
	.
\end{align}
\end{subequations}
With reference to these expressions, and using the notation $\bfR_m=\bfr-\bfr_m$ and $\bfR_m'=\bfr'-\bfr_m$, define the formal expressions
\begin{subequations}
\begin{align}
\delta G_{00}(\bfr,\bfr')=&
	\sum_{mn}
	\biggl(
		T_{00}(\bfR_{mn})\Bigl(g_0(\bfR_m)g_0(\bfR_n')
		-
		\bfg_1(\bfR_m)\cdot\bfg_1(\bfR_n')\Bigr)
		-
		g_0(\bfR_m)\bfT_{11}(\bfR_{mn})\cdot\bfg_1(\bfR_n')
		-
		\bfg_1(\bfR_m)\cdot\bfT_{11}(\bfR_{mn})g_0(\bfR_n')
\nonumber\\&
		-
		i
		\Bigl(\bfg_1(\bfR_m)\times\bfT_{11}(\bfR_{mn})\Bigr)\cdot\bfg_1(\bfR_n')
	\biggr)
	,
\\
\delta G_{01}(\bfr,\bfr')=&
	\sum_{mn}
	\biggl(
		T_{01}(\bfR_{mn})\Bigl(g_0(\bfR_m)g_0(\bfR_n')-\bfg_1(\bfR_m)\cdot\bfg_1(\bfR_n')\Bigr)
		-
		g_0(\bfR_m)\bfT_{10}(\bfR_{mn})\cdot\bfg_1(\bfR_n')
		+
		\bfg_1(\bfR_m)\cdot\bfT_{10}(\bfR_{mn})g_0(\bfR_n')
\nonumber\\&
		-
		i
		\Bigl(\bfg_1(\bfR_m)\times\bfT_{10}(\bfR_{mn}\Bigr)\cdot\bfg_1(\bfR_n')
	\biggr)
	,
\\
\delta\bfG_{10}(\bfr,\bfr')=&
	\sum_{mn}
	\biggl(
		T_{01}(\bfR_{mn})
		\Bigl(
			\bfg_1(\bfR_m)g_0(\bfR_n')
			-
			g_0(\bfR_m)\bfg_1(\bfR_n')
			-
			i\bfg_1(\bfR_m)\times\bfg_1(\bfR_n')
		\Bigr)
		+
		\bfT_{10}
		\Bigl(
			g_0(\bfR_m)g_0(\bfR_n')
			+
			\bfg_1(\bfR_m)\cdot\bfg_1(\bfR_n')
		\Bigr)
\nonumber\\&
		-
		\bfg_1(\bfR_m)\bfT_{10}(\bfR_{mn})\cdot\bfg_1(\bfR_n')
		-
		\bfg_1(\bfR_m)\cdot\bfT_{10}(\bfR_{mn})\bfg_1(\bfR_n')
%\nonumber\\&
		+
		i\bfg_1(\bfR_m)\times\bfT_{10}(\bfR_{mn})g_0(\bfR_n')
		-
		ig_0(\bfR_m)\bfT_{10}(\bfR_{mn})\times\bfg_1(\bfR_n')
	\biggr)
	,
\\
\delta\bfG_{11}(\bfr,\bfr')=&
	\sum_{mn}
	\biggl(
		T_{00}(\bfR_{mn})
		\Bigl(
			\bfg_1(\bfR_m)g_0(\bfR_n')
			-
			g_0(\bfR_m)\bfg_1(\bfR_n')
			-
			i\bfg_1(\bfR_m)\times\bfg_1(\bfR_n')
		\Bigr)
		+
		\bfT_{11}(\bfR_{mn})
		\Bigl(
			g_0(\bfR_m)g_0(\bfR_n')
			+
			\bfg_1(\bfR_m)\cdot\bfg_1(\bfR_n')
		\Bigr)
\nonumber\\&
		-
		\bfg_1(\bfR_m)\bfT_{11}(\bfR_{mn})\cdot\bfg_1(\bfR_n')
		-
		\bfg_1(\bfR_m)\cdot\bfT_{11}(\bfR_{mn})\bfg_1(\bfR_n')
%\nonumber\\&
		+
		i\bfg_1(\bfR_m)\times\bfT_{11}(\bfR_{mn})g_0(\bfR_n')
		-
		ig_0(\bfR_m)\bfT_{11}(\bfR_{mn})\times\bfg_1(\bfR_n')
	\biggr)
	.
\end{align}
\end{subequations}
\end{widetext}

\subsection{Anti-symmetric anisotropy}
In Ref. \cite{fransson_microscopic_2017}, a general expression for the Dzyaloshinskii-Moriya-like interaction $\bfD(\bfr,\bfr')$ between two spins located at $\bfr$ and $\bfr'$, respectively was derived, here repeated for convenience,
\begin{align}
\bfD(\bfr,\bfr')\sim&
	v^2\re
	\int
		f(\omega)
		\Bigl\{
			G_{01}(\bfr,\bfr')\bfG_{10}(\bfr',\bfr)
\nonumber\\&
			+
			G_{00}(\bfr,\bfr')\bfG_{11}(\bfr',\bfr)
		\Bigr\}
	d\omega
	.
\end{align}
In this expression, $v$ denotes the local exchange interaction between the electron spin and the localized spin moment, while $f(\omega)$ is the Fermi-Dirac distribution function. Then, using the example discussed here, it is easy to see that this interaction has a  nonvanishing component also in the absence of $\bfg_1$. Since any kind of spin texture, e.g., spin-polarization or spin-orbit coupling, in the unperturbed electronic structure is accounted for by $\bfg_1$, this implies that there may arise a nonvanishing Dzyaloshinskii-Moriya-like contribution to the spin-spin interactions also for a trivial spin-degenerate electron gas. Indeed, whenever $\bfg_1=0$, the above derivation leads to that $\delta G_{01}(\bfr,\bfr')=0$ and
\begin{align}
\delta\bfG_{11}(\bfr,\bfr')=&
	\sum_{mn}
		g_0(\bfr-\bfr_m)\bfT_{1}(\bfR_{mn})g_0(\bfr_n-\bfr')
\nonumber\\=&
	i\sum_{mn}
		\frac{g_0(\bfr-\bfr_m)g_0(R_{mn})g_0(\bfr_n-\bfr')}{(1-g_0^2|\bfm_m|^2)(1-g_0^2|\bfm_n|^2)}
		\bfm_m\times\bfm_n
	.
\end{align}
Hence, taking $G_{00}\approx g_0$, we obtain
\begin{align}
\bfD(\bfr,\bfr')=&
	\frac{N_0^3}{4\pi}
	\re
	\sum_{mn}
		\bfm_m\times\bfm_n
		\int
			f(\omega)
				H_0^{(1)}(\kappa|\bfr-\bfr'|)
\nonumber\\&
		\times
				\frac{H_0^{(1)}(\kappa|\bfr'-\bfr_m|)H_0^{(1)}(\kappa|\bfr-\bfr_n|)}
					{(1+N_0^2|\bfm_m|^2/2)(1+N_0^2|\bfm_n|^2/4)}
		d\omega
	,
 \label{eq-simDMI}
\end{align}
showing the existence of a nonvanishing Dzyaloshinskii-Moriya interaction between spin moments whenever they are in a noncollinear configuration.
%Moreover, using this example, we notice that this interaction is nonvanishing for collinear spin moments in presence of an intrinsic spin texture. 

In Ref. \cite{JPhysChemLett.14.4941}, chiral molecules were adsorbed onto Cu and Au surfaces, resulting in strongly modified effective spin-orbit coupling and ferromagnetism, respectively, in the two compounds. The magnetic properties associated with the composite structure of chiral molecules interfaced with metals were suggested to result from the type of impurity induced Dzyaloshinskii-Moriya-like interaction formulated in Eq. \eqref{eq-simDMI}.

\section{Decomposition of spin-current Green functions.}
\label{app:B}
As explained in Refs.~\onlinecite{cardias_first-principles_2020}, \onlinecite{cardias_dzyaloshinskii-moriya_2020} and \onlinecite{szilva_quantitative_2022} there exist sum rules which the self-consistent  Green function has to fulfill. Here we sketch the most relevant relations arising from such sum rules for a magnetic dimer on a nonmagnetic substrate, while the general case with more details will be presented in a coming report.  It is possible to divide the terms into two cases depending on whether they have nonrelativistic ($n$) or relativistic ($r$) origin, which is based on  whether they are independent of spin-orbit coupling or not. In addition we conclude that for the dimer case it is only the first term of Eq.~\eqref{defD2} that comes into play, and that in this term it is the factor $\vec{G}^1$ that are most reference dependent.

Firstly, we will make use of a  sum rule for this time reversal even spin dependent, or spin current related, Green function, $\vec{G}^1$, for which  the most important  nonrelativistic and relativistic contributions, respectively, take the form
\begin{align}
	\vec{G}^{1n}_{il}&\approx\sum_{jk}
 %\bigg\{  G^{00}_{ij}\,\lambda_{jk}\,\vec{G}^1_{kl}+\vec{G}^1_{ij}\,\lambda_{jk}\,G^{00}_{kl} 
%	+ \nonumber\\&+
i
 %\left(\vec{G}^1_{ij}\times\,\lambda_{jk}\,\vec{G}^1_{kl}-
 \vec{G}^0_{ij}\times\,\lambda_{jk}\,\vec{G}^0_{kl}
 %\right)\bigg\}
\label{G11-NR}	\\
 \vec{G}^{1r}_{il}&\approx-\sum_{jk}\bigg\{
	G^{00}_{ij}\,\vec{\xi}_{jk}\,G^{00}_{kl}
 + %	\vec{G}^{1}_{ij}\cdot\vec{\xi}_{jk}\,\vec{G}^{1}_{kl}-\vec{G}^{0}_{ij}\cdot\vec{\xi}_{jk}\,\vec{G}^{0}_{kl}+
%	\nonumber\\& %G^{01}\,\vec{\xi}\,G^{00}-G^{00}\,\vec{\xi}\,G^{01}+
	%\nonumber\\&+
%	i\left(\vec{G}^{1}_{ij}\times\vec{\xi}_{jk}\,G^{00}_{kl}+G^{00}_{ij}\,\vec{\xi}_{jk}\times\vec{G}^{1}_{kl}\right)+
%	\nonumber\\
	%&+i\left(\vec{G}^{0}\times\vec{\xi}\,G^{00}-G^{00}\,\vec{\xi}\times\vec{G}^{0}-\vec{G}^{1}\times\vec{\xi}\,G^{01}+G^{01}\,\vec{\xi}\times\vec{G}^{1}\right)+\nonumber\\
%	&%\vec{G}^{0}\cdot\vec{\xi}_{jk}\,\vec{G}^{1}-\vec{G}^{1}\cdot\vec{\xi}_{jk}\,\vec{G}^{0}+
%	\nonumber\\
%	&-%\left\{
%	-\left(\vec{G}^{1}_{ij}\times\vec{\xi}_{jk}\right)\times\vec{G}^{1}_{kl}
%	-\vec{G}^{1}_{ij}\times\left(\vec{\xi}_{jk}\times\vec{G}^{1}_{kl}\right) +
% \nonumber\\
	%-\left(\vec{G}^{1}\times\vec{\xi}_{jk}\right)\times\vec{G}^{0}%+\left(\vec{G}^{0}\times\vec{\xi}_{jk}\right)\times\vec{G}^{1}
%	&+
 \left(\vec{G}^{0}_{ij}\times\vec{\xi}_{jk}\right)\times\vec{G}^{0}_{kl}
%	+\vec{G}^{0}_{ij}\times\left(\vec{\xi}_{jk}\times\vec{G}^{0}_{kl}\right)
\bigg\}\label{G11-R}\,,
\end{align}
with 
\begin{align}
	\lambda_{jk} &=\frac{1}{2}\,\mathrm{Tr}_s \left(G^0_{jk} {}^{-1}+\Delta V_j\delta_{jk}\right)\\
	\vec{\xi}_{jk}&=\frac{1}{2}\,\mathrm{Tr}_s\,\vec{\sigma}\left(G^0_{jk}{}^{-1}+\Delta V_j\delta_{jk}\right),
\end{align}
where the latter is spin dependent due to the fact that it is directly proportional to the spin-orbit coupling.

Since both these terms include $\vec{G}^0$ factors we further observe that the two most relevant terms in its sum rule are 
% time reversal odd and nonrelativistic part of the $\vec{G}^0$ Green function
\begin{align}
	\vec{G}^{0n}_{ik}\approx-\sum_{j}& \bigg\{
 {G}^{00}_{ij}\vec{B}_j{G}^{00}_{jk}-\vec{G}^{0n}_{ij}\vec{B}_j\cdot \vec{G}^{0n}_{jk}
 %+\nonumber\\
% &+iG^{00}_{ij}\left(\vec{B}_j\times\vec{G}^{1n}_{jk}\right)+i\left(\vec{G}^{1n}_{ij}\times\vec{B}_j\right)G^{00}_{jk}+\nonumber\\
%	&+\vec{G}^{0n}_{ij}\times\left(\vec{B}_j\times \vec{G}^{0n}_{jk}\right)-\vec{G}^{1n}_{ij}\times\left(\vec{B}_j\times \vec{G}^{1n}_{jk}\right) 
\bigg\}
\,.\label{G10n}
	\end{align}
 Hitherto, the sum over sites are not confined to two magnetic sites, so these sum rules are relevant also for other multi-site cases.
 
Now we will systematically substitute the $\vec{G}^{0n}$ factors in Eqs.~\eqref{G11-NR} and \eqref{G11-R}
by the terms in Eq.~\eqref{G10n}. In the first iterations this leads for the nonrelativistic case to
\begin{widetext}
\begin{align}
\vec{G}^{1n}_{ij}\approx& i\sum_{kl}\vec{G}^{0n}_{ik}\times\lambda_{kl}\vec{G}^{0n}_{lj}=\nonumber\\
\approx &i\sum_{klmn}\left({G}^{00}_{ik}\vec{B}_k{G}^{00}_{kl}-\vec{G}^{0n}_{ik}
\vec{B}_k\cdot \vec{G}^{0n}_{kl}
\right)\times\lambda_{lm}\left({G}^{00}_{mn}\vec{B}_n{G}^{00}_{nj}
-\vec{G}^{0n}_{lm}\vec{B}_m\cdot \vec{G}^{0n}_{mj}\right)=\nonumber\\
\approx& i\sum_{klmn}{G}^{00}_{ik}\vec{B}_k{G}^{00}_{kl}\times\lambda_{lm}{G}^{00}_{mn}\vec{B}_n{G}^{00}_{nj}+\nonumber\\ 
-&i\sum_{klmnpq}\left\{{G}^{00}_{ik}\vec{B}_k{G}^{00}_{kp}\left(\vec{B}_p{G}^{00}_{pq}\cdot\vec{B}_q{G}^{00}_{ql}\right)\times\lambda_{lm}{G}^{00}_{mn}\vec{B}_n{G}^{00}_{nj}+{G}^{00}_{ik}\vec{B}_k{G}^{00}_{kl}\times\lambda_{lm}{G}^{00}_{mn}\vec{B}_n{G}^{00}_{np}\left(\vec{B}_p{G}^{00}_{pq}\cdot\vec{B}_q{G}^{00}_{qj}\right)\right\}+
\nonumber\\
+&i\sum_{klmnpqrs}{G}^{00}_{ik}\vec{B}_k{G}^{00}_{kp}\left(\vec{B}_p{G}^{00}_{pq}\cdot\vec{B}_q{G}^{00}_{ql}\right)\times\lambda_{lm}{G}^{00}_{mn}\vec{B}_n{G}^{00}_{nr}\left(\vec{B}_r{G}^{00}_{rs}\cdot\vec{B}_s{G}^{00}_{sj}\right)+\ldots
=\nonumber\\
\approx&\left\{A_0+A_1 (\hat{m}_1\cdot\hat{m}_2)+A_2 \left(\hat{m}_1\cdot\hat{m}_2\right)^2\right\}\,(\hat{m}_1\times\hat{m}_2)\label{G11n-expansion},
\end{align}
where only in the last step we have restricted to a dimer. In general, the prefactors $A_i$ are linear combinations of multi-spin products, whose forms are obtained by iterations that identify the nonvanishing multi-spin terms with lowest number of spins, but here with only two spin moments they correspond to higher order interactions of these.

For the relativistic term the same approach gives
\begin{align}
\vec{G}^{1r}_{ij}\approx& -\sum_{kl}G^{00}_{ik}\vec{\xi}_{kl} G^{00}_{lj}-\sum_{klmn}\left\{\left({G}^{00}_{im}\vec{B}_m{G}^{00}_{mk}-\vec{G}^{0n}_{im}
\vec{B}_m\cdot \vec{G}^{0n}_{mk}
\right)\times\vec{\xi}_{kl}\right\}\times\left({G}^{00}_{ln}\vec{B}_n{G}^{00}_{nj}-\vec{G}^{0n}_{ln}
\vec{B}_n\cdot \vec{G}^{0n}_{nj}
\right)=\nonumber\\
\approx& -\sum_{kl}G^{00}_{ik}\vec{\xi}_{kl} G^{00}_{lj}-\sum_{klmn}{G}^{00}_{im}{B}_m{G}^{00}_{mk}\vec{\xi}_{kl}{G}^{00}_{ln}{B}_n{G}^{00}_{nj}+\nonumber\\
+&\sum_{klmnpq}\left\{{G}^{00}_{im}{B}_m{G}^{00}_{mp}\left(\vec{B}_p{G}^{00}_{pq}\cdot\vec{B}_q{G}^{00}_{qk}\right)\vec{\xi}_{kl}{G}^{00}_{ln}{B}_n{G}^{00}_{nj}+{G}^{00}_{im}{B}_m{G}^{00}_{mk}\vec{\xi}_{kl}{G}^{00}_{ln}{B}_n{G}^{00}_{np}\left(\vec{B}_p{G}^{00}_{pq}\cdot\vec{B}_q{G}^{00}_{qj}\right)\right\}+\nonumber\\
-&\sum_{klmnpqrs}{G}^{00}_{im}{B}_m{G}^{00}_{mp}\left(\vec{B}_p{G}^{00}_{pq}\cdot\vec{B}_q{G}^{00}_{qk}\right)\vec{\xi}_{kl}{G}^{00}_{ln}{B}_n{G}^{00}_{nr}\left(\vec{B}_r{G}^{00}_{rs}\cdot\vec{B}_s{G}^{00}_{sj}\right)+\ldots=\nonumber\\
\approx&\left(B_0+B_1(\hat{m}_1\cdot\hat{m}_2)+B_2(\hat{m}_1\cdot\hat{m}_2)^2\right)\,\hat{\xi}\label{G11r-expansion},
\end{align}
\end{widetext}
where the prefactors  $B_i$ are again sums of multi-spin products. 
The site summations of Eqs.~\eqref{G11n-expansion} and \eqref{G11r-expansion} lead uniquely to the final reference dependence only when when there are no intrasite Green functions in the product. Intrasite Green functions instead lead to trivial factors, such as $\hat{m}_1\cdot\hat{m}_1=1$. For instance, the third and fourth terms in Eq.~\eqref{G11r-expansion} contributes to a constant $B_0$ for those terms where $p=q$, but to $B_1$ when $p\neq q$.

Lastly, we  substitute the $\vec{G}^1$ factor in the first term of the Dzyaloshinskii-Moriya interaction in Eq.~\eqref{defD2} by the relations of Eqs.~\eqref{G11n-expansion} and \eqref{G11r-expansion}
and by integrating and summing over the two dimer sites for the factors $A_i$ and $B_i$ we obtain the interactions strength given by $a_i$ and $b_i$ in Eq.~\eqref{eq:multispin1} and  \eqref{eq:multispin2}.

\section{Details of the DFT calculations}
\label{app:C}
The electronic structure calculations were performed using the first-principles, self-consistent real space - linear muffin-tin orbital - atomic sphere approximation (RS-LMTO-ASA) method \cite{frota-pessoa_first-principles_1992,klautau_magnetic_2004,klautau_orbital_2005,frota-pessoa_influence_2002,bezerra-neto_complex_2013,Carvalho2021,Miranda2021,Cardias2016,PhysRevB.96.144413,Ribeiro2011,Igarashi2012b,Kvashnin2016,PhysRevB.97.224402}, which has been generalized to the treatment of noncollinear magnetism \cite{bergman_non-collinear_2006,bergman_magnetic_2007}. This method is based on the LMTO-ASA formalism \cite{andersen_linear_1975},  and solves the eigenvalue problem directly in real space using the Haydock recursion method \cite{haydock_recursive_1980}. 
The local spin density approximation (LSDA) was employed for the exchange-correlation energy \cite{barth_local_1972}. 
To terminate the continued fraction that arises in the recursion method, we used the Beer-Pettifor terminator \cite{beer_recursion_1984} after 21 recursion levels.
 The spin-orbit coupling interaction was included by adding a term  $\xi L \cdot S$  to the Kohn-Sham Hamiltonian, where $\xi$ was calculated self-consistently at each variational step.
 When calculations without spin-orbit coupling are mentioned in the text, it means that $\xi$ was set to zero after the self-consistent calculation.
 
%%%%%%% 
Here, we have considered dimers of Cr, Fe, and Mn supported on Cu(001), Pt(001), and W(001) surfaces. The Cu and Pt substrates were simulated by a cluster containing $\sim 15000$ atoms located in an fcc lattice with the experimental lattice parameter of Cu and Pt, respectively. The W surface was treated analogously using a bcc lattice. 
In order to provide a basis for the wave function near the surface and to handle charge transfers accurately, we included two overlayers of empty spheres above the surface layer. 
The calculations of the dimers were performed by embedding the clusters as a perturbation on the self-consistently converged surfaces. The dimer sites and those of the closest shell of Cu, Pt, W, or empty spheres sites around the defect were recalculated self-consistently, while the electronic structure for all other sites remained unchanged at their clean surface values.

\bibliography{dmiref}
\end{document}